\documentclass[%
 aip,
 amsmath,amssymb,
 reprint,%
]{revtex4-1}

\usepackage{graphicx}
\usepackage{dcolumn}
\usepackage{bm}

\usepackage[utf8]{inputenc}
\usepackage[T1]{fontenc}
\usepackage{mathptmx}
\usepackage{etoolbox}
\usepackage[ruled,vlined]{algorithm2e}
\usepackage{booktabs}

\makeatletter
\def\@email#1#2{%
 \endgroup
 \patchcmd{\titleblock@produce}
  {\frontmatter@RRAPformat}
  {\frontmatter@RRAPformat{\produce@RRAP{*#1\href{mailto:#2}{#2}}}\frontmatter@RRAPformat}
  {}{}
}%
\makeatother

\begin{document}

\preprint{AIP/123-QED}

\title[]{Machine-learning accelerated turbulence modelling \\of transient flashing jets}
\author{David Schmidt}
 \email{schmidt@acad.umass.edu;rmaulik@anl.gov;konstantinos.lyras@kcl.ac.uk.}
  \affiliation{ 
University of Massachusetts, Amherst MA, USA.
}%
\author{Romit Maulik}%
\affiliation{ 
Mathematics and Computer Science Division, Argonne National Laboratory, Lemont, Illinois-60439, USA
}%

\author{Konstantinos G. Lyras}
\affiliation{%
BMEIS, King's College London, London, UK
}%

\date{\today}

\begin{abstract}
Modelling the sudden depressurisation of superheated liquids through nozzles is a challenge because the pressure drop causes rapid flash boiling of the liquid. The resulting jet usually demonstrates a wide range of structures, including ligaments and droplets, due to both mechanical and thermodynamic effects. As the simulation comprises increasingly numerous phenomena, the computational cost begins to increase. One way to moderate the additional cost is to use machine learning surrogacy for specific elements of the calculations. The present study presents a machine learning-assisted computational fluid dynamics approach for simulating the atomisation of flashing liquids accounting for distinct stages, from primary atomisation to secondary break-up to small droplets using the $\Sigma-Y$ model coupled with the homogeneous relaxation model. 
Notably, the model for the thermodynamic non-equilibrium (HRM) and  $\Sigma-Y$ are coupled, for the first time, with a deep neural network that simulates the turbulence quantities, which are then used in the prediction of superheated liquid jet atomisation. 
 The data-driven component of the method is used for turbulence modelling, avoiding the solution of the two-equation turbulence model typically used for Reynolds-averaged Navier–Stokes simulations for these problems. 
Both the accuracy and speed of the hybrid approach are evaluated, demonstrating adequate accuracy and at least 25$\%$ faster computational fluid dynamics simulations than the traditional approach. This acceleration suggests that perhaps additional components of the calculation could be replaced for even further benefit.

\end{abstract}

\maketitle

\section{Introduction}
\begin{figure}
\includegraphics[scale=0.63]{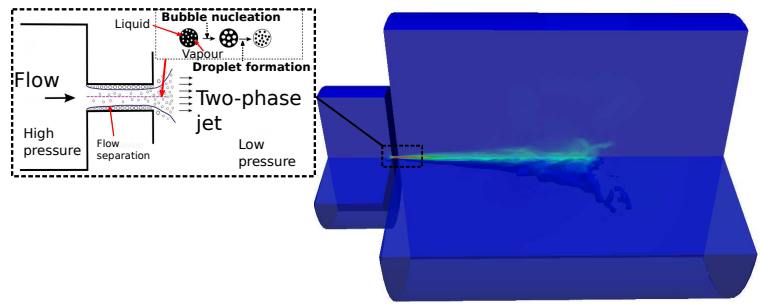}
\caption{\label{flashing_scematics} Schematic of the domain used in simulations. The superheated cryogen flows from a high-pressure region through the nozzle and exits at the low-pressure region as a two-phase jet.}
\end{figure}

Flash boiling is the rapid phase change of a fluid that emerges to ambient conditions where the pressure is below the vapour pressure. Usually, this occurs when a fluid is discharged from a high pressure, high temperature reservoir \cite{Oza1984,Ishii1975}. Much of the early research in flashing discharge was motivated by the need to understand loss of coolant accidents (LOCA) in nuclear power plants due to  accidental releases through cracks in pipes and vessels. Other applications include improving fuel spray atomisation during injection in internal combustion (IC) engines and predicting the flow of cryogenic rocket propellants \cite{Lamanna2015, lyras2021}. In all these cases, the release results in a spray at the nozzle exit which disperses following turbulent mixing, aerodynamic break-up, and droplet collisions \cite{ParkLee1994,YellowBook,Benajes2004,ClearyPhD}. The whole process is not fully understood, but in general the stages into which it is divided are nucleation, bubble growth, and atomisation \cite{YildizPhD,Sher2008}. Flashing can occur either inside or outside the nozzle, depending on the local pressure and geometry, among other factors, leading to interfacial interactions that eventually influence the spray properties \cite{Park1997,Wang2017}.
In strongly flashing conditions, the vapour release results in a spray at the nozzle exit which disperses following turbulent mixing and aerodynamic break-up, producing an explosive two-phase jet of liquid blobs and droplets as shown in Fig.~\ref{flashing_scematics}. 
  
The seminal work in understanding flow rate in short, flashing nozzles is due to Fauske \cite{Fauske1965}, who observed sensitivity to upstream conditions and nozzle length. The impact of flashing on the mass flow rate has later been highlighted by Park et al.\cite{Park1997} who investigated internal flashing for nozzles with different lengths and also concluded that the flashing inception will initiate inside the nozzle. They found that the upstream pressure has a significant impact regardless of the sub-cooling and superheat degree on the mass flow rate through the nozzle. 

Additionally, Park et al.\cite{Park1997} identified two regions inside the nozzle where the pressure drops, one at the inlet and the other at the outlet, where the pressure becomes equal to the ambient value.  
The results were also in agreement with the work of Xu et al.\cite{Xu1995} who studied various values for the stagnation pressure and the degree of sub-cooling showing that the mass flow rate is less sensitive to the latter for short nozzles. On the other hand, stagnation pressure was a major factor for changes in mass flow rate for all the nozzles. 

The type of fluid is also another parameter that may change both the impact on the internal flow and the quality of the atomisation. Simoneau\cite{Simoneau1975}, in their experiments for flashing liquid nitrogen, concluded that a metastable jet is present which leads to a two-phase jet right at the orifice exit. 
Similarly, results were reported in Hendricks et al.\cite{Hendricks1976} for flashing liquid oxygen through converging-diverging nozzles under high pressure. 

Usually, the result of the rapid pressure drop associated with flashing leads to bubble nucleation and changes the flow to an extent which, apart from the pressure, depends on the ratio of the nozzle length $L$ to its diameter $D$ \cite{YildizPhD,Sher2008}. Yildiz\cite{YildizPhD} has shown that higher $L/D$ increases bubble nucleation and leads to more vapour in the flow. Wang et al.\cite{Wang2017} in their experiments for nozzles with high $L/D$ observed that bubble nucleation occurs in random positions for flashing R134 flowing inside the channel at either a region close to the inlet or along the whole nozzle and closer to the exit with increasing pressure. A unified theory that helps to explain these behaviours using fundamental thermodynamics has been presented by Lammana et al. \cite{lamanna2014}.
 
Computational fluid dynamics (CFD) models in the literature have been developed in an attempt to successfully simulate inter-phase heat transfer and mass flow under choking conditions which are likely to occur in flashing flow \cite{lyras2021, dang2018, karathanassis2017, gartner2020}. In such situations, there are two limiting cases:  infinitely fast phase change and infinitely slow phase change. These extremes can be represented by the Homogeneous Equilibrium Model (HEM) or the frozen flow model, respectively. The work of Fauske \cite{Fauske1965} indicates that neither limiting behaviour is adequate for simulating flash boiling in short nozzles. Consequently, flash-boiling models necessarily contain some finite-rate heat or mass transfer expression.

The Homogeneous Relaxation Model (HRM), correlated by Downar-Zapolski et al.\cite{DownarZapolski1996}, is based on the theoretical approach presented by Bilicki and Kestin \cite{Bilicki1990} and is  able to capture heat transfer under flashing conditions accounting for the non-equilibrium vapour generation. The idea is to consider a relaxation term such that the instantaneous quality would relax to the equilibrium value over a locally varying timescale. Although it is possible to consider this time-scale constant, here it is calculated via an empirical correlation. The HRM is implemented by adding this relaxation term to the transport equation for vapour generation.  
Schmidt et al. \cite{Schmidt2010} have proposed an algorithm which links the standard pressure-velocity coupling algorithm to the HRM, developing a compressible solver which was tested for the internal flashing of water for various tests. Based on this work, Lyras et al.\cite{Lyras2019numerical} presented a unified approach for modelling both the internal flow and liquid atomisation that linked the involved processes such as evaporation, turbulent mixing to the algorithm for calculating pressure. The method employed the Eulerian-Lagrangian-Spray-Atomisation (ELSA) model for atomisation and has been validated for various tests for flashing cryogenic liquids and water \cite{LyrasIJMF}. 

The flash-boiling simulations are an example of a typical multi-phenomena simulation where several different sub-models are invoked for representing their respective phenomenon.  With the inclusion of each additional physical phenomenon (flashing, atomisation, turbulence) the number of equations and thus the computational cost increases.  In this paper, the method of Lyras et al.\cite{Lyras2019numerical} is accelerated by substituting part of the simulation by a deep neural network.
Specifically, a machine learning surrogate model is used for completely substituting the turbulence model used in previous works in Lyras et al.\cite{Lyras2019numerical}. Recently, deep learning and other data-driven methods for turbulence closure modelling have shown promise in accelerating LES \cite{maulik2019subgrid,maulik2018data,maulik2020spatiotemporally,yuan2020deconvolutional,xie2020modeling,shin2021data,xie2019artificial,wang2021artificial,yao2020modeling,subel2021data} and RANS simulations \cite{ling2015evaluation,ling2016machine,tracey2015machine,ling2016reynolds,cruz2019use,yang2020improving}. The basic premise of the vast majority of such models is that high fidelity direct numerical simulation data is used to learn optimal subgrid or Reynolds stress closures. This could be through a direct learning of the stresses, learning the unresolved force vectors (i.e., the divergence of the stress), or through learning a projection on a suitable space. These may also be through some form of outer-loop intelligence such as reinforcement learning \cite{novati2021automating} or gene expression programming for symbolic regression \cite{weatheritt2016novel,zhao2020rans,beetham2020formulating} or by neural network in loop adjoint optimisation \cite{sirignano2020dpm}. The reader is directed to the excellent review of \citet{duraisamy2020perspectives} for an in-depth discussion of recent advances in data-driven closure modelling.

In some studies, quantities of interest (in the absence of training data) are used to calibrate these data-driven closures. In contrast, a few studies have looked at data-driven turbulence closure modelling from the perspective of building surrogates for currently available models. For example Zhu et al.\cite{zhu2019machine} and Maulik et al.\cite{maulik2019accelerating} built surrogates for the Spalart-Allmaras model \cite{allmaras2012modifications} for parameterised airfoil and backward facing step geometries. The formulation presented in the latter showed a significant acceleration (up to 5-7X) when compared to several one and two-equation-based models and their utilisation for steady-state incompressible flows. 

The present effort builds on Maulik et al.\cite{maulik2019accelerating} by evaluating this surrogate turbulence modelling approach for a two-equation model but for \emph{transient flows} relevant to flash boiling.  We note the key fact that such types of problems may easily be parameterised and the deployment of an accurate surrogate model could reduce the simulation cost significantly. Therefore, the proposed hybrid machine learning-CFD approach aims to substitute the turbulence model computations in the transient solver with a graph created by a trained neural network which could additionally speed up the solver. The primary goal of the paper here is to validate the hybrid approach and compare the obtained simulation times with CFD. The parts of the solution algorithm are analysed first and the hybrid solver is tested for nozzles with different geometry. We observe that our results are in good agreement with both experimental data and classical numerical computations. Moreover, we obtain a 25$\%$ reduction in execution time suggesting that the solver can be used for successfully simulating the metastable jet both inside and outside the nozzle.

\section{Numerical modelling of flash boiling}\label{sec:parent_model}
\subsection{Non-equilibrium vapour generation}
The mass fraction for vapour $x$ is calculated solving the following transport equation 

\begin{equation} 
\frac{\partial \rho x}{\partial t} + \frac{\partial \rho u_{j} x}{\partial x_{j}}=\Gamma
\label{ref:x-equation}
\end{equation} 
Here, $\rho$ is the mixture density which is the weighted sum of the liquid and vapour densities and $u_{j}$ is the velocity field. The source term $\Gamma$ corresponds to the vapour generation rate and is modelled with the HRM \cite{DownarZapolski1996}.

Assuming that $x$ relaxes towards an equilibrium value $x_{eq}$ over a time-scale $\Theta$ the HRM is states that, 
\begin{equation} 
\Gamma = -\rho \left(\frac{x - x_{eq}}{\Theta} \right)
\label{ref:Gamma-equation}
\end{equation}
The boiling process following an abrupt pressure drop is a non-equilibrium vapour generation process \cite{Sher2008}. This implies that although saturation conditions are reached the liquid undergoes a metastable state and will not start boiling instantaneously. This time delay is represented in Eq.~\ref{ref:x-equation} with the time-scale $\Theta$. The equilibrium vapour mass fraction is calculated as, 
\begin{equation} 
x_{eq}= \frac{h-h_{l,sat}}{h_{v,sat}-h_{l,sat}}
\label{ref:xbar-equation}
\end{equation} 
where $h_{l,sat}, h_{v,sat}$ are the saturated enthalpies of the liquid and vapour state. 

Based on experiments for flashing water and for pressure up to 10 bar in Downar-Zapolski et al.\cite{DownarZapolski1996} the time-scale $\Theta$ for the model is calculated as, 
\begin{equation} 
\Theta = \Theta_{0}\alpha^{-0.257}\psi^{-2.24}
\label{ref:Theta-equation}
\end{equation} 
The non-dimensional pressure $\psi$ is equal to $\left(p_{sat}-p\right)/p_{sat}$. The time-scale $\Theta_{0}$ is a constant of the model, $\Theta_{0}=6.51\times 10^{-4} [s]$ and the void fraction $\alpha$ is calculated from the local densities as, 

\begin{equation} 
\alpha = \frac{\rho_{l} - \rho}{\rho_{l}-\rho_{v}}
\label{ref:VFrac-equation}
\end{equation}           
where $\rho_{l}$ and $\rho_{v}$ are the liquid and vapour densities. The dimensionless pressure denoted with $\psi$ is defined as, 
\begin{equation} 
\psi = \left| \frac{p_{sat} - p}{p_{sat}} \right|
\label{ref:pdimless_less_than_10bar-equation}
\end{equation} 
For higher pressures above $10\: bar$ the following expression has been proposed for the relaxation time, 
\begin{equation}
\Theta = \Theta_{0}\alpha^{-0.54}\psi^{-1.76}
\end{equation}
The constants of the model in this case become $\Theta_{0}=3.84 \cdot 10^{7} [s]$ and 
\begin{equation} 
\psi = \left| \frac{p_{sat} - p}{p_{crit}-p_{sat}}\right|
\label{ref:pdimless-equation}
\end{equation}  
The value of $\Theta$ tends to decrease with increasing void fraction and non-dimensional pressure \cite{DownarZapolski1996,Schmidt2010}.
The HRM expressions have been used before for superheated R134A and liquid nitrogen \cite{LyrasIlass,LyrasIJCMEM} and are known to work for other superheated liquids as well such as iso-octane \cite{Moulai2015} and JP8 \cite{Lee2009}. The model has been shown to perform well without adjustment of constants either in the present work or in former efforts \cite{Moulai2015, Schmidt2010, baldwin2016, Lee2009}.

\subsection{Pressure-velocity coupling}
The HRM and the presented method here are implemented in the in-house multiphase code based on the open source CFD code OpenFOAM \cite{Weller1998}. A segregated approach is used for solving pressure, velocity, density and other equations. 

An additional equation for considering the effects of air entrainment to the mixture is solved here. 
The scalar $y$ denotes the mass fraction of the vaporising species (liquid and its vapour) which means that $y$ = 1 is for no air and $y$ = 0 in the case of only non-condensible gas (air) in a computational cell. The equation for $y$ is
\begin{equation}
\frac{\partial \rho y}{\partial t}+ \frac{\partial \rho u_{j} y}{\partial x_{j}} =
\frac{\partial}{\partial x_{j}} \left(\frac{\mu_{t}}{Sc_{t}} \frac{\partial y}{\partial x_{j}} \right)
\end{equation}
where $u_{j}, \rho , \nu_{t}, and Sc_{t}$ are the velocity in the $j$ direction, the mixture density, the kinematic turbulent viscosity and the turbulent Schmidt number respectively. After $y$ is advected the liquid mass fraction is then calculated as,

\begin{equation}
y_l = \frac{y(1-x)(\rho_{l}-(\rho_{l}-\rho_{v}) x) }{\rho_{l}}
\end{equation}

The momentum equation can be written in a matrix notation following Jasak \cite{JasakPhD} as
\begin{equation} 
a_{P}u_{P} = H(u_{j})- \frac{\partial p}{\partial x_{i}} + F_{\sigma}
\label{ref:momentum-equation}
\end{equation}
Here, $a_{P}$ is the matrix that contains the diagonal coefficients for a cell $P$ and $H(u_{j})$ is the coefficient matrix for all the neighbouring cells of $P$ with any other source terms apart from the pressure gradient and $F_{\sigma}$ is the surface tension force. 

Bilicki and Kestin\cite{Bilicki1990} proposed the following expression for the material derivative of density with respect to pressure, temperature and vapour quality, $\rho=\rho(p,h,x)$,
\begin{equation} 
\frac{D\rho}{Dt} = \left(\frac{\partial \rho}{\partial p}\right)_{h,x} \frac{Dp}{Dt} + \left(\frac{\partial \rho}{\partial h}\right)_{p,x} \frac{Dh}{Dt} + \left(\frac{\partial \rho}{\partial x} \right)_{p,h}\frac{Dx}{Dt}
\label{ref:material_derivative-equation}
\end{equation}

Due to the continuity equation, the left-hand-side of Eq.\eqref{ref:material_derivative-equation} becomes
$-\rho\frac{\partial u_{i}}{\partial x_{i}}$. Solving for $u_{P}$ in Eq.\eqref{ref:momentum-equation} and substituting into Eq.\eqref{ref:material_derivative-equation} an equation for pressure is derived. The matrix form of the pressure and velocity denoted with $u$ reads, 

\begin{equation} 
\begin{aligned}
& \rho\frac{\partial}{\partial x_{j}}\left(\frac{1}{a_{p}}H(u_{j})\right)_{f}-\rho\frac{\partial}{\partial x_{j}}\left(\frac{1}{a_{p}} \frac{\partial p}{\partial x_{i}}\right)+\rho\frac{\partial}{\partial x_{j}} \left(\frac{1}{a_{p}}F_{\sigma} \right) \\
& + \left(\frac{\partial \rho}{\partial p}\right)_{h,x} \frac{Dp}{Dt} + \left(\frac{\partial \rho}{\partial h}\right)_{p,x} \frac{Dh}{Dt}  
+\left(\frac{\partial \rho}{\partial x}\right)_{p,h}\frac{Dx}{Dt}=0
\end{aligned}
\label{ref:post_pressure-equation}
\end{equation}
Substituting the HRM expression into the pressure equation becomes 
\begin{equation} 
\begin{aligned}
& \rho\frac{\partial}{\partial x_{j}}\left(\frac{1}{a_{p}}H(u_{j})\right)_{f}-\rho\frac{\partial}{\partial x_{j}}\left(\frac{1}{a_{p}} \frac{\partial p}{\partial x_{i}}\right)+\rho\frac{\partial}{\partial x_{j}} \left(\frac{1}{a_{p}}F_{\sigma} \right) \\
& + \left(\frac{\partial \rho}{\partial p}\right)_{h,x} \frac{Dp}{Dt} + \left(\frac{\partial \rho}{\partial h}\right)_{p,x} \frac{Dh}{Dt} +\left(\frac{\partial \rho}{\partial x}\right)_{p,h}\left(\frac{x-x_{eq}}{\Theta}\right) \\
& +\left(\frac{\partial \rho}{\partial y}\right)\frac{Dy}{Dt} =0
\end{aligned}
\label{ref:pressure-equation}
\end{equation}
The above equation considers the effect of the involved processes of thermal non-equilibrium and multiphase mixing and the surface tension force for the generated spray. 

The variables were stored in the cell centres in a co-located arrangement and are interpolated at the cell faces. The convective terms here were discretised with a second-order bounded scheme that provides accuracy and stability as shown in Jasak et al.\cite{Jasak1999} which blends the upwind (low order) and central (second order) schemes and offers a smooth transition between the two. For the gradient terms, a second order scheme with a linear correction is used.  
For the predictor/corrector cycle, 3 PIMPLE loops were used for updating the matrix which contains all of the terms the momentum equation, except for the gradient of pressure.
This matrix is then updated and used for calculating the fluxes without the contribution of $\nabla p$. 
For the inner cycles for pressure, 5 to 10 PISO loops were used for solving for the pressure 
and updating the velocity field using Courant numbers up to 2. 
At the inlet, fixed values were imposed for pressure boundary conditions, and for the outlet a boundary condition developed by Poinsot and Lele\cite{Poinsot1992} was employed. At the outlet, zero gradients were assumed for velocity. 

\section{Liquid atomisation}\label{sec:liquidAtomisation}

The high Weber number, high Reynolds number atomisation process created by flashing nozzles leads to a wide range of length and time scales in the ensuing two-phase flow. For most applications, resolving the interfacial details is impractically expensive. Analogous to using a turbulence model for unresolved inertial effects, we apply an interfacial evolution model for unresolved interfacial features. The present approach avoids assumptions of specific droplet morphologies, instead employing a more general representation of the interface.

For modelling the spray, the $\Sigma$-Y model model is used here for a fixed-grid Eulerian framework. To define the liquid ligaments and droplets, the approach solves an equation for the evolution of the surface density $\Sigma$ in both space and time. 
The latter is expressed in units $[1/m]$ and denotes the amount of liquid/gas interface inside a differential volume as a result of the various processes during atomisation that can cause either surface generation or destruction \cite{Vallet1999,Vallet2001}.

In this paper the model proposed by Lyras et al.\cite{Lyras2019numerical} is implemented in the solver as an extra equation to solve within the segregated algorithm. 
The model reads,
\begin{equation} 
\begin{aligned}
& \frac{\partial \bar{\Sigma}}{\partial t} + \frac{\partial \tilde{u_{j}} \bar {\Sigma}}{\partial x_{j}} =                  
   \frac{\partial}{\partial x_{j}}\left(\frac{\nu_{t}}{Sc_{t}}\frac{\partial \bar {\Sigma}}{\partial   x_{j}}\right)+\Psi\left(S_{init} + S_{turb}+S_{vap,den} \right)+ \\
& \left(1-\Psi \right) \left(S_{coll} + S_{2ndBU} + S_{vap,dil}\right) 
\end{aligned}
\label{ref:4.53-equation}
\end{equation}
Here, different source terms are organised based on whether they concern the primary atomisation region or the dilute region.
In the first case, the terms describe the surface changes due to turbulence ($S_{turb}$) and evaporation ($S_{vap,den}$), whereas in the dilute region these include the contributions of the droplets collisions ($S_{coll}$), the aerodynamic break-up ($S_{2ndBU}$) and vaporisation ($S_{vap,dil}$). 
The Reynolds average, $\bar{\Sigma}$ is used here for the Reynolds-averaged Navier–Stokes equations (RANS) simulations and $\tilde{u_{j}}$ is the mass weighted Favre average of velocity. 
The phase indicator $\Psi$ is 1 if the liquid mass fraction, $\tilde{Y_{l}}$ is between 0.5 and 1, and is zero for cells with a liquid mass fraction less than 0.1. For all other cases $\Psi$ is calculated as in Lyras et al.\cite{LyrasIlass} from the liquid volume fraction, $\phi_{l}$, where $\phi_{l}=\bar{\rho}\tilde{Y_{l}}/\bar{\rho_{l}}$ as,

\begin{eqnarray}
\Psi(\phi_{l})=H(\phi_{l}-0.1)H(\phi_{l}-0.5) \nonumber\\
+(H(\phi_{l}-0.1)-H(\phi_{l}-0.5))(2.5\phi_{l}-0.25)
\label{4.34-equation}
\end{eqnarray}
where $H()$ is the Heaviside step function. The terms, apart from $S_{init}$, in Eq.\eqref{ref:4.53-equation} are generally of the following form, 
\begin{equation}
S=\frac{\bar{\Sigma}}{\tau_{\Sigma}}\left( 1- \frac{\bar{\Sigma}}{\bar{\Sigma}_{eq}} \right)
\label{4.35-equation}
\end{equation}
where $\bar{\Sigma}_{eq},\tau_{\Sigma}$ are an equilibrium value for the interface and the time-scale of the corresponding process \cite{Vallet1999}. 
The term $S_{init}$ is related to the minimum amount of liquid-gas surface produced during atomisation process which is inversely proportional to the characteristic turbulent spatial scale $l_t$ obtained here by the k-$\omega$-SST model. The expression from Menard et al.\cite{Menard2006} reads

\begin{equation}
S_{init}=Y_l(1-Y_l)/l_t
\end{equation}

With small values of liquid mass fraction this becomes 
\begin{equation}
S_{init}= 2\frac{\mu_{t}}{Sc_{t}} \frac{6 \bar{\rho}}{\rho_{l}\rho_{g}l_{t}}\frac{\partial \tilde{Y_{l}}}{\partial x_{i}} \frac{\partial \tilde{Y_{l}}}{\partial x_{i}}
\end{equation}

where $\rho_{l},\rho_{g}$ are liquid and gas densities. 
Analytical expressions and details for the source terms $S_{init}, S_{turb}, S_{coll}, S_{2nBU}, S_{vap,den}$ and  $S_{vap,dil}$ are provided in Lyras et al. \cite{Lyras2019numerical} providing expressions specifically for superheated jets for the evaporation effects. 
 
After solving the surface density equation, the Sauter Mean Diameter (SMD) (the $D_{32}$) can be 
estimated by the expression
\begin{equation}
D_{32} = \frac{6 \bar{\Sigma}}{\tilde{Y_{l}}}
\label{4.50-equation}
\end{equation}

\section{Machine learning in PIMPLE}
For the RANS simulations in the present work, a machine learning surrogate model is used which is similar to that developed in Maulik et al.\cite{maulik2019accelerating}. In this approach the turbulent viscosity is not calculated by solving the equations for turbulence modelling, but are predicted by a surrogate turbulence model. Specifically, for steady-state cases, a data-driven prediction would be utilised to obtain the converged turbulent eddy-viscosity following which the momentum and pressure equations would be solved to convergence numerically. It was observed that large relaxation factors could be used in the steady state solver to accelerate the time to solution for steady state test cases. 

In contrast, the segregated solver in the present work uses the PIMPLE algorithm, which is a combination of the Pressure implicit with splitting of operator (PISO) \cite{Issa1986} and Semi-Implicit Method for Pressure-Linked Equations (SIMPLE)\cite{Patankar1980} algorithms.  More importantly, the turbulent eddy-viscosity flow field is \emph{transient} and therefore, the data-driven model requires evaluation at each time step of the numerical solver. We emphasize that the core idea of this surrogate turbulence model is the deployment of a computational graph that predicts turbulent viscosity without solving the equations for any approximate model. In the present paper, we replace the k-$\omega$-SST model, which is comprised of two equations. These networks are used for validation as well as training data generation. These equations are not solved during the PISO/SIMPLE iterations, but instead the turbulent viscosity is predicted based on the trained parameters of the ML model. The general framework of the ML-surrogate model deployment within the PIMPLE algorithm is shown below.

\vspace{0.5cm}

\begin{algorithm}[H]
\SetAlgoLined
\KwResult{Updates $p-U$}
 Load ML graph\;
 \While{Predictor/Corrector}{
  Solve $x,y,\rho$\;
  ML surrogate model (nothing to solve) or turbulence model $\rightarrow$ $\nu_{t}$ \;
  \While{PISO}{
   $U$ without $p$-correction\;
   Volume flux calculation\;
   Solve for $\rho_l, \rho_g, \bar{x}, h$ from EOS or NIST \;
   Calculate D$\rho$/dt, D$x$/dt,D$h$/dt,$\partial \rho / \partial p$, $\partial \rho /\partial x$, $\partial \rho /\partial h$, $\partial \rho /\partial y$ $\rightarrow$ D$\rho$/dp\;
   Solve $p$-equation\;
   Update $U$ with $p$-correction\; 
   }
   Compute mass fluxes\;
 } 
 \caption{Pressure-velocity coupling algorithm}
\end{algorithm}

\vspace{0.5cm}

The obtained $\nu_{t}$ is used in the equations for the liquid vapour mass fraction, the mixture $y$, and the surface density among others. The successful construction and deployment of a surrogate comprises the data generation phase, the training phase, and finally the online deployment of the generated computational graph within PIMPLE. First, some simulations are preformed to obtain the data required for the ML model. 

Although the results shown here are primarily for validation and an initial assessment of the capabilities of the proposed transient surrogate, there is great potential for generalising the results. This is because flash boiling significantly depends on the $L/D$, $D$, the degree of superheat of the jet, the initial pressure, and the type of fluid which may all be parameters that affect potential surrogate models. Considering all the above makes it possible to predict the flow regime inside the nozzle and the atomisation and spray characteristics for a wide range of downstream conditions \cite{Rachakonda2018}. This is particularly important due to the scarce available experiments for evaluating the characteristics of pressurised jets of fluids used in various industries and energy production, such as cryogenic liquids for instance, liquefied natural gas (LNG), liquid hydrogen and others.

The second phase for the ML surrogate model implementation is the training of the neural network based on the generated data. The paper here uses a simple fully-connected architecture defined in Tensorflow. The training is done using the Python API, and the trained model is saved to disk to be loaded by OpenFOAM during run time\footnote{https://github.com/argonne-lcf/TensorFlowFoam}. The inputs to the neural network are the diameter of the nozzle, the local velocity components, the local coordinates, and the mixture marker function. The latter was found to significantly improve training and the overall accuracy of the surrogate model. The output of the network is given by the turbulent eddy viscosity. Note that we vary the diameter of the nozzle (and consequently the mass flow rate) to create various training and test cases. Training cases are used to fit the surrogate model whereas unseen test cases are used for validation of the constructed surrogate. The implementation details of the trained model  to work alongside the OpenFOAM libraries for turbulence and the analysis for the neural network architecture are provided in Maulik et al.\cite{maulik2019accelerating}. 

Various different set-ups were tested, and a network with 3-5 hidden layers and 20-40 neurons were found to give fast results with adequate accuracy for this approach (denoted as ML-CFD hereafter) when compared to standard CFD. Although the neural network here is used to avoid the equations for the turbulence model (here for $k$ and $\omega$), this work could be easily extended to avoid other equations: for instance for the liquid vapour mass fraction, the mixture $y$, and the surface density, among others. More trainable parameters can be easily added to the graph to make the online deployment possible for these variables, thus reducing the number of equations considered in the solution algorithm.  

\section{Results and discussion}
Results for assessing the method are organised as follows: First some results from numerical simulations of flashing R-404a cryogen are shown. Experimental data are also presented from literature for evaluating the capability of the CFD solver and the accuracy. The second test consists of simulations performed with the ML-CFD technique in two-dimensional jets of water and its comparison with the classical CFD simulations. The last results presented concern mass flow rate calculated for different initial pressures for flashing jets of water. The results include comparisons with other works in order to assess relative accuracy. 

\subsection{Validation with 3D R404a jets}
In this test the R-404a cryogen is saturated and stored at a pressure of 1.25 MPa ~\citet{zhou2012experimental}. The flow in the nozzle considered in the test is controlled through a valve which is placed between the storage vessel and the nozzle. Once the valve is open, the fluid flows towards the exit of the nozzle connected with the valve downstream the storage vessel. The conditions at the nozzle exit are the normal atmospheric conditions. Once the jet is released, the result is a two-phase spray which fragments into smaller ligaments which break-up during their motion in the ambient relatively hot environment. The geometry of the nozzle used for the experiment is a stainless steel tube of length $L=63.5$ mm with inner diameter $D=0.81$ mm.
For validating our solver, the corresponding simulations used a mesh with 2 million hexahedral cells.  The extent with the nozzle exit and the farfield domain was $55D$ in the radial direction and $200D$ at the jet axis so that the impact of the boundary conditions would be minimised. 

\begin{figure}
\includegraphics[scale=0.3]{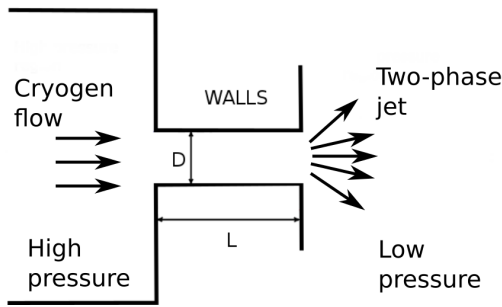}
\caption{\label{figdomain} Schematic of the domain typically used in the present simulations. The pure liquid flows from a high-pressure region through the nozzle and exits at the low-pressure region as a liquid-gas mixture in the form of a jet.}
\end{figure}

The domain used for simulations is shown in Fig.~\ref{figdomain}. All the physical parameters of the experiment are listed in Table ~\ref{tab:Zhifu}. 

\begin{table}
\caption{\label{tab:Zhifu}Physical properties for the 3D R404a simulations.  }
\begin{ruledtabular}
\begin{tabular}{lcr}
Physical parameter for simulations & Value \\
\hline
Inlet pressure & 1.25 $MPa$ \\
Outlet pressure & 100 $kPa$ \\
Outlet temperature & 298 $K$ \\
$L/D$ & 78.4  \\
Nozzle diameter & 0.81mm    \\
Thermodynamic conditions & Saturated \\

\end{tabular}
\end{ruledtabular}
\end{table}

\begin{figure}
\includegraphics[scale=0.135]{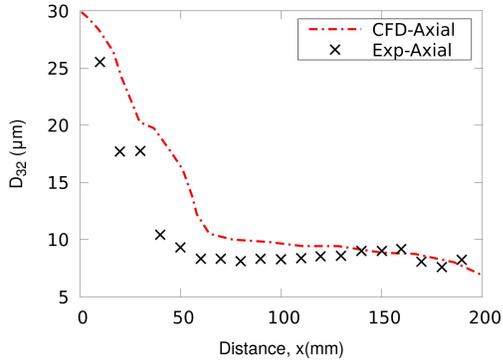}
\caption{\label{r404a_d32} Sauter mean diameter ($D_{32}$) at the centreline. Comparison with Zhou et al.\cite{zhou2012experimental}.}
\end{figure}

Fig.~\ref{r404a_d32} shows the results for the average droplet Sauter-Mean-Diameter ($D_{32}$) along the spray axis of the spray.
There are two regions that describe distinct evolution patterns for the droplets: the region for the nozzle exit up to $x=50mm$ and another one from $x=50mm$ up to the end of the domain. 
At the first region, the primary atomisation region, larger droplets emerging from the nozzle become smaller in size with a steep gradient in the $D_{32}$ graph. 
Based on experiments, the regime of at the end of the nozzle is most likely to be a two-phase mixture with the presence of vapour phase in the form of bubbles. The rapid depressurisation results in bubble nucleation which starts inside the nozzle.

The effect of the $L/D$, the storage conditions, and the nozzle diameter  directly influence the bubble formation and its impact on the atomisation process. Henry \cite{Henry1970} reported that for $L/D$ more than 5 the liquid jet is expected to break up into smaller droplets with the formation of vapour bubbles at the centreline of the two-phase jet. 
This was also observed here, in both the simulation and the experiment. 
In the experiment, droplets with temperature below the boiling point were observed, suggesting that
bubbles have formed inside the nozzle. The most important term of Eq.\eqref{ref:4.53-equation} in this region is $S_{init}$ which is proportional to the gradient of the liquid mass fraction as described in the equation for the surface density. As a consequence, this gradient promotes an increase in surface density.

After a distance of approximately 50mm the droplet diameter stops decreasing with the increasing distance from the nozzle and becomes relatively constant. 
This is expected since the jet disperses in all three directions drastically decreasing the diameter of the larger droplets. 
In terms of modelling, this is pronounced with a decrease in the liquid volume fraction. The liquid volume fraction is maximum at the nozzle exit and starts to progressively drop until it becomes small enough to consider it negligible. It has been shown before that the rate of change for the volume fraction of the liquid phase is much greater than the surface density which, depending the position, it might be orders of magnitude larger than the volume fraction (it is inversely proportional to $D_{32}$)\cite{Lyras2019numerical}. The terms responsible for the secondary break-up of the spray in the $\Sigma$-equation are more important in the dilute region of the spray. Particle collisions, aerodynamic break-up, and evaporation significantly result in the fragmentation of the liquid ligaments and blobs to smaller pieces which spread in the farfield until the size of the produced droplets becomes uniform. 

Although the CFD results provide reasonable predictions for the Sauter mean diameter, the results are subject to improvements related to the submodels used here. Previous experiments have shown that the size of the generated droplets undergoing flash-boiling atomisation depend on pressure and the type of the fluid\cite{Zhifu2012}. Here, the HRM model parameters are used with the original values in Downar-Zapolski et al.\cite{DownarZapolski1996} for water. Although the accuracy of the results here is adequate, a machine-learning technique could be built based on a graph-like system for specifying the appropriate vapour fraction $x$ and avoid solving Eq.\eqref{ref:x-equation} entirely.

\begin{figure}
\includegraphics[scale=0.135]{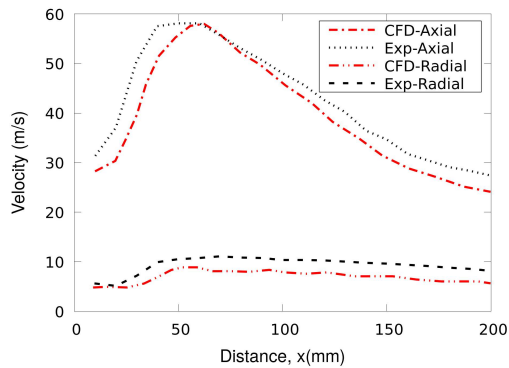}
\caption{\label{fig404a_velocity} Axial and radial velocity along the centre line. Comparison with Zhou et al.\cite{zhou2012experimental}.}
\end{figure}

The results for the axial and radial velocity are shown in Fig.~\ref{fig404a_velocity}. The velocity at the jet axis is much higher than the radial velocity, and the jet maintains a liquid core which is much denser at the centreline. 
The axial velocity starts increasing at the so-called expansion region until it reaches a maximum at approximately the same distance that marks the large gradient in the $D_{32}$ graph (x=40mm). After that, the velocity starts decreasing in the entrainment region of the spray where the two-phase regime is more evident with smaller liquid blobs. 
The experiments have revealed that the diameter significantly decreases in the axial direction following the internal flashing and the primary atomisation process right after the end of the nozzle.
The acceleration in the primary atomisation region is in agreement with other experiments as in \cite{aguilar2001theoretical,Zhifu2012} and observations in Polanco et al.\cite{Polanco2010}.
 
The droplet radial velocity remained below 10m/s, which is much smaller compared to the axial velocity which had a maximum value of approximately 60m/s. 
Due to the explosive nature of flashing, as a consequence of the flashing inception in the nozzle, the enhanced atomisation of the liquid core at the nozzle exit was the major reason for the acceleration of the droplets. 
At the entrainment region, the size of the droplets decreases both due to aerodynamic break-up and evaporation and becomes at least three times smaller than the size at the exit of the nozzle (approximately at 8$\mu$m). The result is a vapour cloud with small droplets of uniform size dispersed with velocity which gradually decreases. 

\subsection{Further validation and testing with R134A}

This test examined the machine-learning acceleration for simulations of R134A (1,1,1,2 -- Tetrafluoroethane: CF3-CH2F) jets flowing through short sharp nozzles. 
The domain of the problem is similar to the previous problem but with a shorter nozzle. This test was utilised for the ML-CFD assessments in contrast to the previous one which was used solely for the validation of the numerical solver. The computational domain consists of a storage vessel with high pressure where R134A is stored at saturated conditions (700 $kPa$ at 247 $K$). The liquid flows through the nozzle of length $L=4mm$ and diameter $D=3mm$ and is dispersed into the low-pressure region with atmospheric temperature and pressure. The physical properties for the simulations are shown in Table ~\ref{tab:ParkLee1994}.  The tests considered for studying flash-boiling experiments consisted of a straight channel with a sharp inlet.  The problem was represented as an axisymmetric, two-dimensional domain, which is a common assumption for this type of flow problem \cite{Moulai2015, Rachakonda2018}.  

\begin{figure}
\includegraphics[scale=0.135]{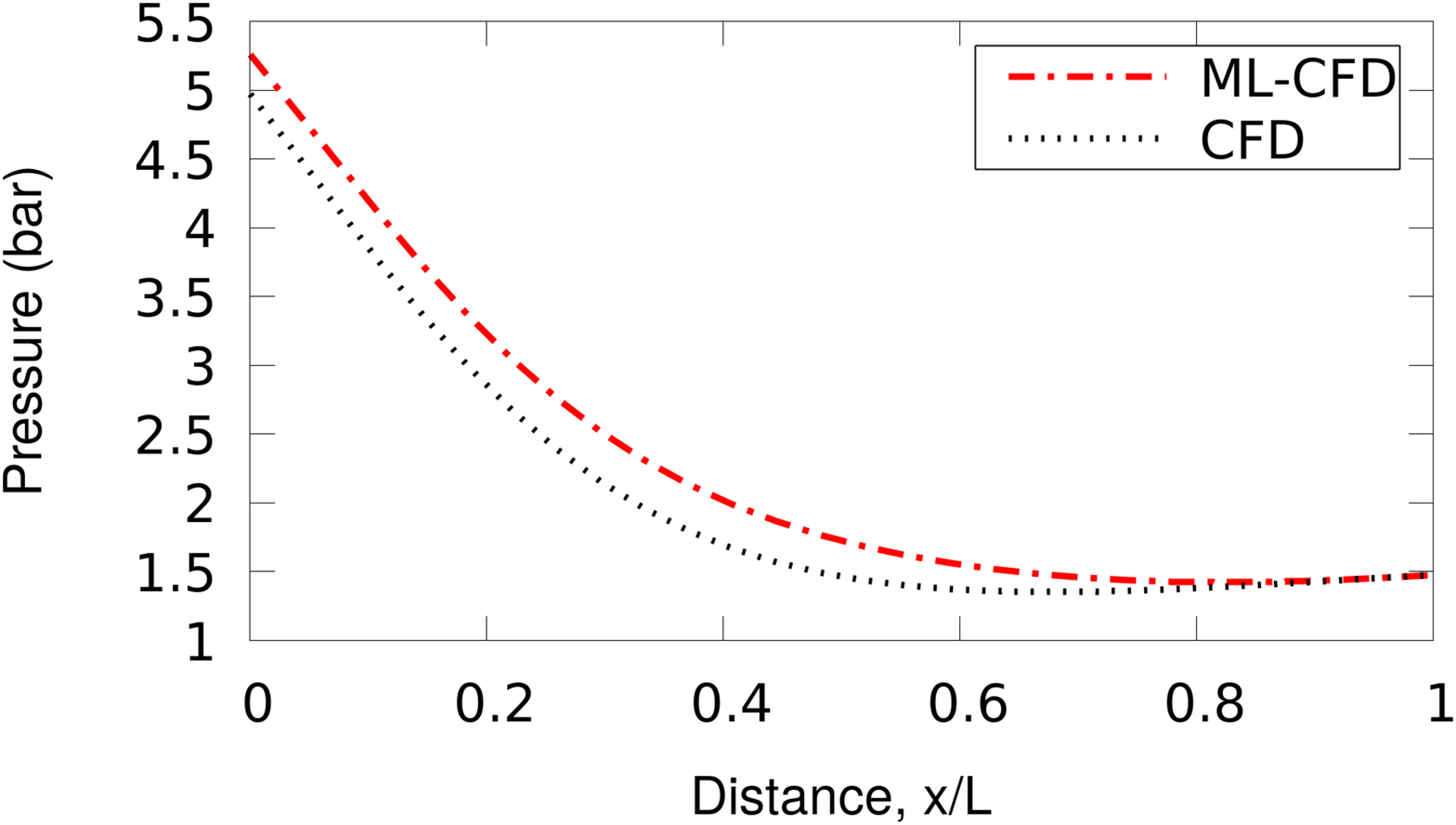}
\caption{\label{2D_r134_116k_p_centreline} Pressure obtained with CFD and ML-CFD at the centreline inside the nozzle with respect to distance for the 2D case.}
\end{figure}

\begin{figure}
\includegraphics[scale=0.135]{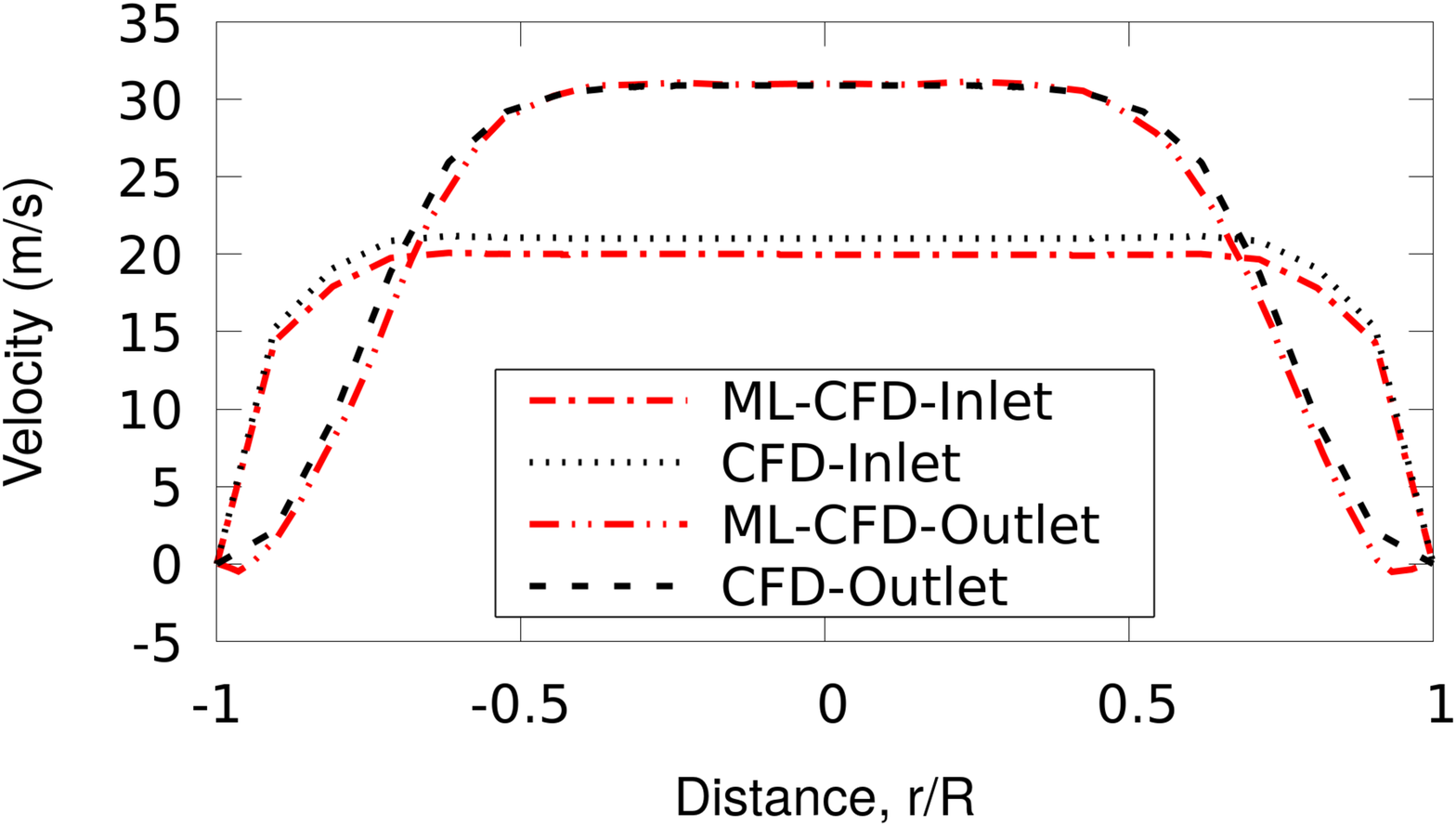}
\caption{\label{2D_r134_116k_ux} Velocity obtained with CFD and ML-CFD at the inlet and the outlet inside the nozzle with respect to distance r/R (R=radius of nozzle) from the walls for the 2D case.}
\end{figure}

Two simulations with 140000 cells were used for simulating two distinct cases: One for $D=2mm$ and another one for $D=4mm$. The results were stored up to the first 0.005 seconds (physical time) of the simulation and the data were used for training the neural network with 20 neurons and 4 hidden layers.  
The input given for the neural network consisted of the velocity components, the coordinates in two dimensions, the vapour mass fraction $x$, and the turbulent viscosity $\nu_{t}$. We have found that including $x$ in the network inputs improves the training, reducing the loss error, and consequently increases the accuracy for the CFD solver.   
The training in Tensorflow for all tests here was conducted using a personal workstation with a E5-2620 Intel Xeon processor.
Depending on the number of neurons and hidden layers, the training may vary for this infrastructure from some minutes up to a day. For the above settings, the training in our tests lasted approximately 3 hours. The next step after training the neural network was simulating the case of $D=3mm$ with keeping all the other physical parameters the same. This case of $D=3mm$ corresponds to a completely new case, different than the data used for training the network. For both classical CFD and ML-CFD, a mesh of 60000 hexahedral cells was used for the simulations. Since the internal flow at the nozzle is of particular importance for flash boiling and its impact on the atomisation process, the flow patterns inside the nozzle were studied comparing pressure and velocity. The pressure along the nozzle centreline is shown in Fig.~\ref{2D_r134_116k_p_centreline}, and in Fig.~\ref{2D_r134_116k_ux} the velocities at both the beginning and the end of the nozzle are presented. In general the ML-CFD results for both pressure and velocity are in good agreement with the CFD ones. 
In Fig.~\ref{2D_r134_116k_p_centreline} pressure decreases in an almost linear fashion, and after x/L=0.6 up to the downstream position (x/L=1) becomes almost constant.
This drop is expected for flashing flows and has been also previously observed\cite{Park1997,Winklhofer2001}.

\begin{figure}
\includegraphics[scale=0.3]{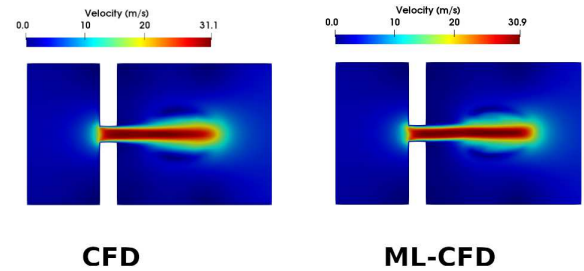}
\caption{\label{2D_r134a_ux_caption} Caption of jet velocity for the selected 2D case. Comparison with CFD and ML-CFD.}
\end{figure}

Although the rapid phase change due to an abrupt pressure drop is more pronounced in longer nozzles with $L/D$ in excess of 10 \cite{Park1997,Sher2008}, bubble nucleation in short nozzles is also possible \cite{YildizPhD,Simoneau1975,Hendricks1976} and also depends on the initial thermodynamic state of the flowing fluid.
Here, an annular flow regime was observed suggesting the existence of vapour pockets in the near-wall region.
This is in agreement with the experiments for other cryogenic fluids flowing through nozzles with small $L/D$ as in Simoneau et al.\cite{Simoneau1975} where the measurements for flashing liquid nitrogen revealed that vaporisation may happen inside the nozzle, changing the jet morphology.  
As shown in Fig.~\ref{2D_r134a_ux_caption} for the velocity at the axial direction, both ML-CFD and CFD approaches produced similar results for velocity. 
In Fig.~\ref{2D_r134_116k_ux} the axial velocity is shown at the upstream and downstream orifices. 
The upstream velocities indicate a more uniform flow and have the maximum value at the centreline of the jet. The velocity increases as expected at the nozzle exit with a less uniform, more parabolic shape of the jet with a maximum of 30m/s which is 30$\%$ higher than in the upstream position. 
Overall, the results follow similar trends as in Lyras et al.\cite{LyrasIJMF}.  

\begin{figure}
\includegraphics[scale=0.73]{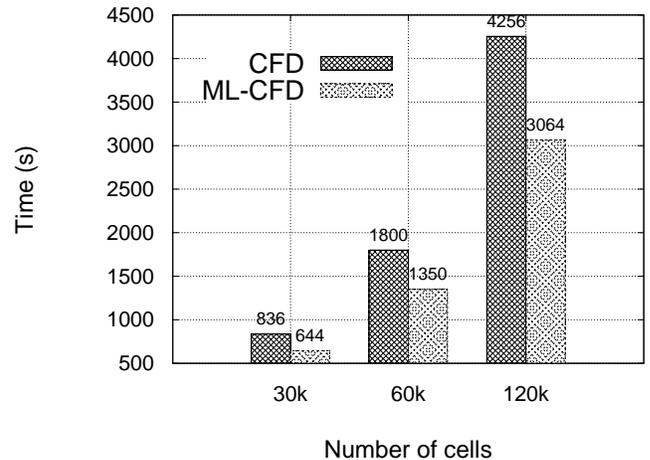}
\caption{\label{2D_r134a_time} CPU time of the 2D R134a case for the different meshes considered.}
\end{figure}

In Fig.~\ref{2D_r134a_time} the execution times are shown for different mesh resolutions for 30000, 60000, and 120000 hexahedral cells. The computational cost difference between the CFD and the ML-CFD approach varied from 26-32$\%$.  
As described before, this is a test that describes the potential of using this hybrid approach in the PISO/SIMPLE algorithms. With a substitution of the turbulence modelling one can see the overall gain in time for the simulations. For the tests here, all meshes were coarser or had the same number of cells than the cases used for training the network. All tests used the same training. 

\begin{table}
\caption{\label{tab:ParkLee1994}Physical properties for the 2D simulations with R134A.  }
\begin{ruledtabular}
\begin{tabular}{lcr}
Physical parameter for simulations & Value \\
\hline
Inlet pressure & 700 $kPa$ \\
Inlet temperature & 247$K$ \\
Outlet pressure & 100 $kPa$ \\
Outlet temperature & 298$K$ \\
$L/D$ & 1.3  \\
Nozzle diameter & 3mm    \\
Thermodynamic state & Saturated \\

\end{tabular}
\end{ruledtabular}
\end{table}

\subsection{Validation and ML testing with water in 2D channels}
The next test is presented to show the capability of the method to calculate the mass flux.
The ML-CFD approach was tested for the experiments of Fauske\cite{Fauske1965} considering flashing water flowing through a tube with $L/D=4$ and $D= 6.35 mm$ with various different inlet pressures for discharging saturated water. A computational domain similar to the one for the prior tests was chosen to simulate scenarios with various initial pressures of 1.37 MPa, 4.13 MPa, and 6.89 MPa. For the cases here, for each one of the pressures considered, two additional simulations were carried out with the same $L$ but with three times the diameter and five times the diameter of the first three calculations. These 6 additional cases were use to generate data for training the 20 neuron neural network with 4 layers. 
For convenience, the flows here were simulated as axisymmetric flow where more changes appear primarily at the axial and radial direction. As before, in addition to the nozzle, a plenum is added at the inlet and at the outlet side of the nozzle. The meshes used for these cases had approximately 138000 hexahedral cells. The results for simulating the case $L/D=4$ are shown in Fig.~\ref{axisym_mass_flux} for the mass flow rate. For the tests, three different meshes with 25000 (coarse), 50000 (medium) and 100000 (fine) cells are shown as a function of the upstream pressure. The results are compared with the experiment and the results from Schmidt et al.\cite{Schmidt2010} demonstrating good agreement with both. All tests used the same trained network.

In the tests here the water flows inside a sharp nozzle and accelerates, reaching its peak velocity at a short distance from the beginning of the nozzle. The stream diameter reaches a minimum (vena contracta) and flow separation is observed here closer to the walls of the nozzle, which is expected \cite{Schmidt2010}. The decrease in pressure right at the nozzle inlet increases the rate of phase change. The vena contracta downstream of the inlet occurs with a smaller drop (compared to the pressure drop at the nozzle exit). Pressure, as shown in the HRM equation, is a significant factor for the vapour generation inside the nozzle. Even though a relatively short nozzle is used for these tests, bubble nucleation occurs and the metastable jet inside the nozzle undergoes phase change right at the tip of the inlet corner, changing its regime at the exit to a vaporised flashing jet.

In the absence of any impurities, as assumed here, during the sudden pressure drop of the fluid with and the acceleration of the flow, the liquid is likely to reach the homogeneous nucleation limit of superheat prior to any significant changes towards a two-phase flow regime.
The vapour bubbles form and grow, not only in proximity to the nozzle walls, but in the liquid core of the jet \cite{Levy2010}. The role of pressure and temperature in forming the vapour that leaves the nozzle wall and its association with the calculated mass flow rate is discussed in detail in Lyras et al. \cite{Lyras2019numerical}.

The grid independence of the ML-CFD results was investigated, as shown in Fig.~\ref{axisym_mass_flux}. The coarse and fine mesh results have demonstrated an acceptably low sensitivity to the grid resolution. This also indicates that the training of ML on a specific mesh may be generalised to another coarse mesh for a similar problem.

The reduction of the simulation time of the CFD and ML-CFD and for the three different meshes as shown in Fig.~\ref{axisym_mass_flux_time} varied from 26$\%$ with the coarse mesh to 32.5$\%$ with the finest mesh. The differences between the simulation times for this case were the same as in the previous test with the shorter nozzle with the lower pressure. In this case, the pressure varied and was much higher. The flow is also different since more vapour is generated as a consequence of the higher $L/D$. 

\begin{figure}
\includegraphics[scale=0.135]{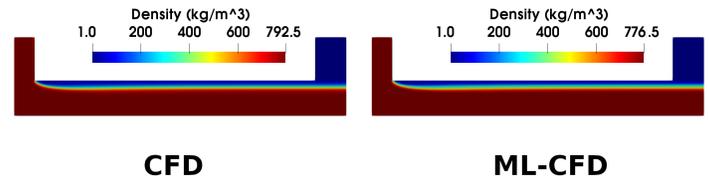}
\caption{\label{axisym_density_caption} Calculated density with CFD and ML-CFD for the experiment of Fauske\cite{Fauske1965} with 1.38 MPa saturated liquid discharge. Since axisymmetry is assumed, the bottom edge of each figure represents the axis.}
\end{figure}

\begin{figure}
\includegraphics[scale=0.135]{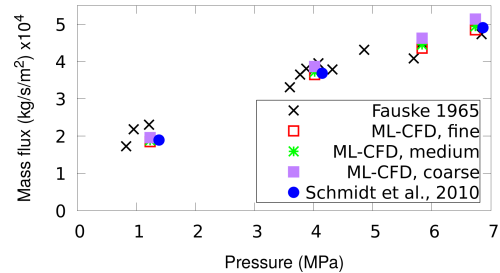}
\caption{\label{axisym_mass_flux} Calculated mass flow rates with ML-CFD for a nozzle with L/D = 4 compared with the calculations from Schmidt et al.\cite{Schmidt2010} and the experimental data from Fauske\cite{Fauske1965}.}
\end{figure}

\begin{figure}
\includegraphics[scale=0.73]{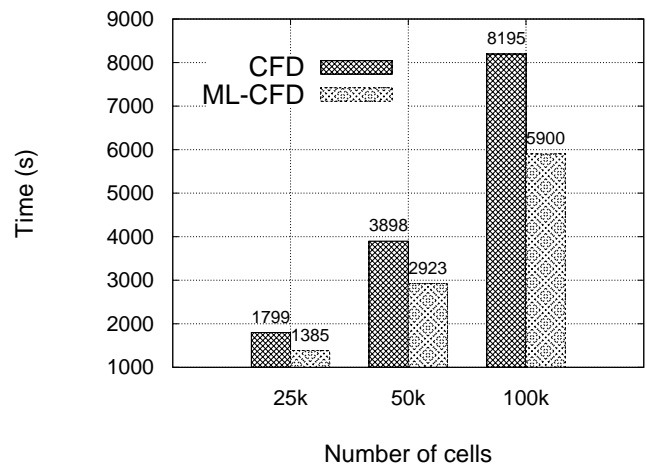}
\caption{\label{axisym_mass_flux_time} CPU time of the 2D flashing water case for the different meshes considered.}
\end{figure}

\section{Conclusion}
In this research, a novel hybrid machine learning-CFD model for simulating the atomisation of flashing jets is presented and validated. The method considers the non-equilibrium vapour generation as a consequence of the small-scale heat and mass transfer phenomena.
The impact of the superheat degree and orifice geometry on the mass flow rate is also examined. The results reveal that accurate calculations can be obtained for the mass flow rates, pressure, and other characteristics and can predict the properties of atomisation in realistic accidental releases. 

The ML-CFD approach achieves at least $25\%$ faster results when compared to the standalone CFD. This is the first approach that couples the ELSA approach and HRM with neural networks. Most notably, the training in the offline phase includes transient simulations without considering steady-state simulations as similar hybrid approaches.
The online phase deploys the graph and updates the turbulent viscosity on the fly.
In addition to these benefits, the ML-CFD approach which links the pressure-velocity coupling algorithm to the neural network can be easily extended to account for more equations. In particular, the vapour mass fraction, the mixture marker function equations, and the equations for enthalpy and density could be completely substituted by a similar surrogate modelling strategy.

Although this paper focuses primarily on the validation of the solver with the machine learning turbulence modelling, the promising results for the simulation time give rise to a series of prospects for even faster simulations. The first step would be replacing the equations for the vapour mass fraction $x$ and the mixture equation $y$. Since the pressure equation involves various terms that encapsulate the underpinning physics, the ultimate goal would be replacing it completely, providing a fast prediction of pressure for the specific problem. The pressure equation is by far the most time-consuming part of the solution algorithm here, and could be replaced by a trained neural network which would consider variables such as $x,y,U$.  These potential expansions of the current approach are the subjects of subsequent investigations.

\begin{acknowledgments}
The submitted manuscript has been created by UChicago Argonne, LLC, Operator of Argonne National Laboratory (Argonne). Argonne, a U.S. Department of Energy (DOE) Office of Science laboratory, is operated under Contract No. DE-AC02-06CH11357. This research also used resources of the Argonne Leadership Computing Facility, which is a DOE Office of Science User Facility supported under Contract DE-AC02-06CH11357. The U.S. Government retains for itself, and others acting on its behalf, a paid-up nonexclusive, irrevocable worldwide license in said article to reproduce, prepare derivative works, distribute copies to the public, and perform publicly and display publicly, by or on behalf of the Government.

The authors declare that they have no conflict of interest.
\end{acknowledgments}

\section*{Data Availability Statement}
The data that support the findings of this study are available from the corresponding author upon reasonable request.

\bibliography{aipsamp}

\end{document}